\documentclass{elsart}

\usepackage{epsfig}
\usepackage{amsmath}

\usepackage{graphicx}

\input amssym

\begin{document}

\begin{frontmatter}



\title{A simple evolution equation for rapidity distributions in nucleus-nucleus collisions}

\author{J.~Dias de Deus and J.~G.~Milhano}

\address{CENTRA, Instituto Superior T\'ecnico (IST), Av. Rovisco Pais, P-1049-001 Lisboa, Portugal}

\begin{abstract}
We explore the relationship between the Glasma framework and the String Percolation Model by introducing a simple model for the rapidity distributions in nucleus-nucleus collisions. The model is solved for both symmetrical and asymmetrical collisions. The phenomenon of limiting fragmentation is briefly discussed.
\end{abstract}


\end{frontmatter}

\section{Introduction} \label{sec:intro}

The String Percolation Model (SPM) \cite{Armesto:1996kt} for high energy hadronic collisions has a long history of successfully  describing experimental observations \cite{Pajares:2004ib,DiasdeDeus:2003ei,Pajares:2005kk,DiasdeDeus:2005yt}.
As with any efficacious phenomenological model, one would wish, if not to fully derive it from first principles, to comprehensively understand its relationship  with more fundamentally motivated approaches.

The Glasma, a highly coherent configuration of classical longitudinal chromo-electric and chromo-magnetic fields, has been proposed \cite{Lappi:2006fp,Romatschke:2005pm,Romatschke:2006nk,Lappi:2006nx,Lappi:2006hq,Kharzeev:2006zm,Gelis:2006ks} as an adequate description of the matter formed immediately after a high energy hadronic collision.
Further, it may provide \cite{Romatschke:2005pm,Romatschke:2006nk,Lappi:2006nx,Kharzeev:2006zm,Gelis:2006ks} the necessary physical mechanism for the rapid thermalization observed at RHIC and thus for the eventual creation of the Quark Gluon Plasma.
The Glasma hypothesis follows from the Colour Glass Condensate (CGC) approach to high energy hadronic processes \cite{McLerran:1993ni,McLerran:1993ka,McLerran:1994vd,Jalilian-Marian:1996xn,Jalilian-Marian:1997jx,Jalilian-Marian:1997gr,Jalilian-Marian:1997dw,Kovner:1999bj,Kovner:2000pt,Weigert:2000gi,Iancu:2000hn,Ferreiro:2001qy,Weigert:2005us}, which in turn provides a theoretically sound and phenomenologically successful description for the infinite momentum wavefunction of a hadronic object.
The CGC/Glasma framework tackles the same physical issues as those addressed by the SPM. The relationship between the CGC/Glasma and the SPM has been hinted at repeatedly, but it is only recently that efforts have been made to understand and explore such a connection \cite{Kharzeev:2006zm,DiasdeDeus:2006xk,Armesto:2006bv}.

In the CGC/Glasma framework, a nucleus-nucleus collision, viewed in the reference frame of the centre of mass of the collision, is to be thought of as that of two Lorentz contracted to the transverse plane sheets of coloured glass \cite{Kovner:1995ts,Kovner:1995ja,Krasnitz:1998ns,Krasnitz:1999wc,Krasnitz:2000gz,Krasnitz:2001qu,Lappi:2003bi}.
The CGC description of the infinite momentum hadronic wavefunction relies on the introduction of a longitudinal momentum separation scale.
The small-$x$ component of the wavefunction, a system of non-perturbatively large colour fields, can be described in terms of a classical Weizs\"acker-Williams field radiated by the large-$x$ components of the wavefunction.
Such an effective description, in which the relevant dynamical degrees of freedom (small-$x$ gluons) are described by a classical field with the large-$x$ constituents playing the role of a  classical static (in light-cone time) source, holds for modes with longitudinal momentum close to the separation scale. The separation scale is, however, arbitrary and can be lowered in successive steps towards increasingly smaller $x$. This renormalization group procedure is encoded in the B-JIMWLK evolution equations \cite{Jalilian-Marian:1996xn,Jalilian-Marian:1997jx,Jalilian-Marian:1997gr,Jalilian-Marian:1997dw,Kovner:1999bj,Kovner:2000pt,Weigert:2000gi,Iancu:2000hn,Ferreiro:2001qy,Balitsky:1995ub,Balitsky:1998ya,Mueller:2001uk,Blaizot:2002xy} for the CGC. 
The small-$x$ part of the wavefunction is characterized by a transverse momentum scale --- the saturation momentum $Q_s$ --- arising from finite density induced non-linear interactions between the wee partons (gluons). 
Modes with transverse momentum below $Q_s$ are saturated and cannot be further populated. Any additional gluon emission will thus result in an increase of the saturation momentum.

Immediately after the collision, the hadron-hadron system is described by the highly coherent state of the two CGC sheets. 
The original transverse fields, associated with the large-$x$ components of the hadronic wavefunctions, remain intact and, as the sheets pass through each other, add to themselves a distribution of colour electric and colour magnetic charge.
These charge densities, which  are of equal magnitude but opposite sign, act as sources for longitudinal fields  living in the region between the parting CGC sheets. 

The physical situation described by the Glasma, the system of purely longitudinal fields in the region between the parting hadrons,  is analogous to that underlying the SPM. 
As the effective number of strings, including percolation effects, is directly related to the rapidity particle density, we thus have the `equations'
\[ [\mbox{Glasma}] \sim [\mbox{Effective Strings}] \sim [\mbox{Particle density}] \, .\]
It can thus be argued that once the Glasma is created through the mechanism outlined above and carefully detailed in \cite{Lappi:2006fp,Gelis:2006ks}, its evolution can be phenomenologically described in the framework of the SPM.
In this paper we follow such a line of thought and address the question of the rapidity evolution of a system of longitudinal fields/strings.

In Sec.~\ref{sec:rapitidity} we introduce our simple model and compute the density distribution in rapidity for both symmetrical and asymmetrical nucleus-nucleus collisions. In Sec.~\ref{sec:limfrag} we address the issue of limiting fragmentation. Our conclusions are presented in Sec.~\ref{sec:conc}.

\section{Rapidity distributions}
\label{sec:rapitidity}

\subsection{The model}

The simplest model that can be imagined to describe the generation of lower centre of mass rapidity objects from higher rapidity ones with asymptotic saturation is the nineteenth century logistic equation for the dynamics of populations \cite{logistic1,logistic2}
\begin{equation}\label{eq:log}
	\frac{\partial \rho}{\partial (-\Delta)} = \frac{1}{\delta} (\rho - A \rho^2)\, ,
\end{equation}
where $\rho \equiv \rho (\Delta, Y)$ is the particle density, $Y$ is the beam rapidity, and 
\begin{equation}
	\Delta \equiv |y| - Y\, .
\end{equation}
The variable $-\Delta$ plays the role of evolution time. 
The parameter $\delta$ controls the low density $\rho$ evolution and must therefore depend on intrinsic parameters of the theory. The parameter A, responsible for saturation, will naturally depend on the nature of the target and on the energy $e^Y$.

The $Y$-dependent limiting value of $\rho$, determined by the saturation condition $\frac{\partial \rho}{\partial (-\Delta)} = 0$, is given by 
\begin{equation}\label{eq:rhos}
	\rho_Y = \frac{1}{A}\, .
\end{equation}

The condition $\frac{\partial^2 \rho}{\partial (-\Delta)^2}\big |_{\Delta_0} = 0$, defining the separation between the region $\Delta > \Delta_0$ of low density and positive curvature and the region $\Delta < \Delta_0$ of high density and negative curvature, gives
\begin{equation}\label{eq:rho0}
	\rho_0 \equiv \rho( \Delta_0, Y) =  \frac{\rho_Y}{2}\, .
\end{equation}

\begin{figure}[h] 
   \centering
   \includegraphics[angle=0,width=10cm]{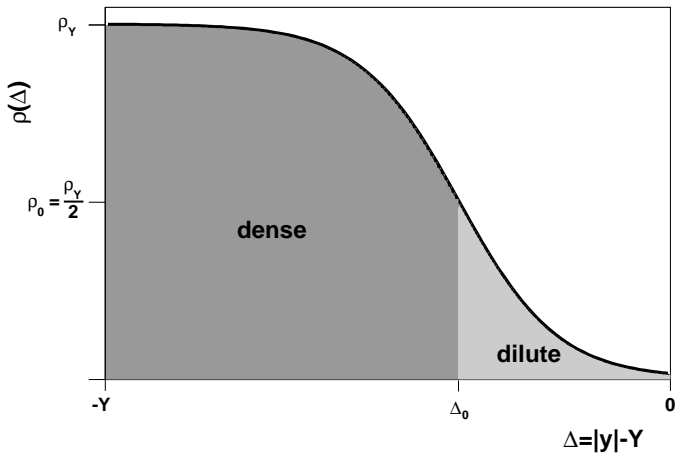} 
   \caption{The distribution (\ref{eq:rhoint}) showing saturation at large $-\Delta$, the change of curvature at the point $\Delta_0$ which separates between the dense ($\Delta < \Delta_0$) and the dilute ($\Delta > \Delta_0$) regions.}
   \label{fig:solution}
\end{figure}

Integrating (\ref{eq:log}) between $\rho_0$ and some $\rho (\Delta)$, and using (\ref{eq:rhos}) and (\ref{eq:rho0}), we obtain
\begin{equation}\label{eq:rhoint}
	\rho( \Delta, Y) = \frac{\rho_Y}{e^{\frac{\Delta - \Delta_0}{\delta}}+1}\, .
\end{equation}
Eq. (\ref{eq:rhoint}), illustrated in Fig.\,\ref{fig:solution}, is nothing but a generalization of the Fermi distribution, and is known to fit RHIC data for rapidity distributions in the case of central nucleus-nucleus colisions \cite{Adams:2005cy,Brogueira:2006nz}.
There are three parameters in (\ref{eq:rhoint}): two from the original differential equation, $\rho_Y$ and $\delta$, and an additional one, $\Delta_0$, from the integration of (\ref{eq:log}).
The parameters $\rho_Y$ and $\Delta_0$ are expected to be energy (or $Y$) dependent, but $\delta$ can be taken, at least approximately, as a constant. In the real Fermi distribution one has  $\rho_Y=1$, but here we are not dealing with a dominantly fermionic system. 

The separation, at $\Delta_0$, between a dilute region ($\Delta > \Delta_0$) of objects with rapidity close to the beam rapidity and a dense region ($\Delta < \Delta_0$) of low rapidity objects is a close analogue of that underlying the CGC framework. 
There, one distinguishes between a dilute set of objects (partons) carrying a large fraction of the hadronic wave function total longitudinal momentum --- the source --- and a dense region of partons with small longitudinal momentum. 
A further dense-dilute separation, that set by the saturation momentum $Q_s$, is considered in the CGC. Such a separation is absent from our simple model since all dependence on tranverse dimensions has been neglected.

\subsection{Symmetrical nucleus-nucleus collision}

So far we have concentrated on following a single nucleon pair. In a nucleus-nucleus collision there is a number $N_{part}$ of nucleons that take part in the collision. In the  symmetrical situation of a $AA$ collision we have 
\begin{equation}
	N_{part} = 2 N_A\, ,
\end{equation}
where $N_A$ is the average number of participants from each nucleus. Both in the CGC approach and in the SPM the particle density is proportional to $N_A$ and we can write
\begin{equation}\label{eq:rhonucleus}
	\rho( \Delta, Y, N_A) =  \frac{N_A \,\rho_Y}{e^{\frac{\Delta - \Delta_0}{\delta}}+1}\, .
\end{equation}

In order to justify (\ref{eq:rhonucleus}) we shall look in more detail at the SPM. The particle density is directly related  to the effective number of strings 
\begin{equation}\label{eq:rhoperc}
	\rho(\Delta) \equiv \frac{d n}{dy} \sim N_s^{eff} \sim F(\eta) N_s(\Delta)\, ,
\end{equation}
where $N_s (\Delta)$ is the number of strings contributing in a given $\Delta$, $F(\eta)$ is the colour summation reduction factor \cite{Braun:2001us}
\begin{equation}
	\label{eq:feta}
	 F(\eta)\equiv \sqrt{\frac{1-e^{-\eta}}{\eta}}\, ,
\end{equation}
and $\eta$ is the percolation transverse density parameter
\begin{equation}
	 \eta \equiv \big(\frac{r}{R}\big)^2 N_s(\Delta)\, .
\end{equation}
Here,  $r$ is the transverse radius of the string and $R$ is the radius of the overlapping area which can be  obtained from simple geometrical arguments
\begin{equation}
	R\simeq R_p N_A^{1/3}\, ,	 
\end{equation}
where $R_p$ is the proton radius.

The percolation transition occurs for \cite{Rodrigues:1998it} $\eta = \eta_c \simeq (1.15  - 1.5)$.

We distinguish two regimes:

\renewcommand{\theenumi}{\roman{enumi}}
\renewcommand{\labelenumi}{\theenumi)}
\begin{enumerate}
\item \textbf{High density regime} [$\Delta < \Delta_0, \, \eta\equiv \eta_{N_A N_A} > \eta_c$]\\
In the dense regime physics is dominated by gluons and sea quarks and the number of strings contributing is proportional to the number of collisions \cite{DiasdeDeus:1997dj,DiasdeDeus:2000cg,DiasdeDeus:2000gf}, that is  
\begin{equation}
	N_s \sim N_A^{4/3}\, . 
\end{equation}	
Further, $\eta$ is large such that  $F(\eta) \rightarrow \eta^{-1/2}$ and (see (\ref{eq:feta})): 
\begin{equation}
	  F(\eta) \sim \frac{1}{N_A^{1/3}}\, ,
\end{equation}
so that  we obtain
\begin{equation}
	  \rho(\Delta) \sim N_A\, .
\end{equation}

\item \textbf{Low density regime} [$\Delta > \Delta_0, \, \eta\equiv \eta_{pp}<\eta_c$]\\
In the dilute region physics is dominated by the valence strings and thus $N_s \sim N_A$, $F(\eta)$ is independent of $N_A$ and we obtain \cite{DiasdeDeus:2000cg,DiasdeDeus:2000gf,Bialas:1976ed}
\begin{equation}
	 \rho(\Delta) \sim N_A\, .
\end{equation}
\end{enumerate}

We find that, at least approximately, the relation $\rho \sim N_A$ holds in both limits thus justifying (\ref{eq:rhonucleus}).

\subsection{Asymmetrical collision}

\begin{figure}[h] 
   \centering
   \includegraphics[width=5cm]{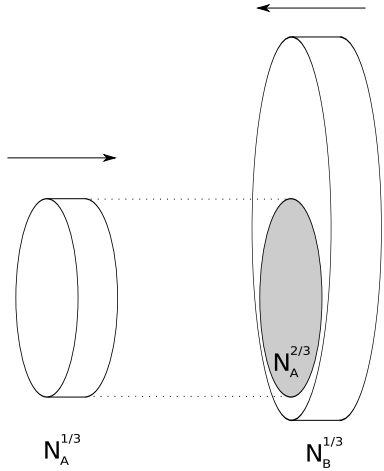} 
   \caption{Estimate of number of collisions for an asymmetrical AB collision (see text).}
   \label{fig:collision}
\end{figure}

We consider next the asymmetrical situation of  $AB$ collisions. In what follows we shall always assume that $A\leqslant B$ and that $A$ moves from left to right. Let us start by making a simple estimate, based on geometrical arguments, of the number $N_{coll}$ of nucleon-nucleon collisions resulting from a given  number of participant nucleons (see Fig.\,\ref{fig:collision}). If a tube of $N_A^{1/3}$ nucleons from nucleus $A$ interacts with  $N_B^{1/3}$ nucleons from nucleus $B$ along the collision axis then, as the overlapping  area of interaction is $N_A^{2/3}$, we have 
\begin{equation}\label{eq:ncoll}
	N_{coll} \simeq \big(N_A^{1/3} N_B^{1/3} \big) N_A^{2/3} = N_A N_B^{1/3} \, .
\end{equation}
Clearly, the well known result from Glauber calculus $N_{coll} \simeq N_A^{4/3}$ is recovered for a symmetrical situation.

The quantities $N_A^{1/3}$ and $N_B^{1/3}$ depend on impact parameter: for $b\simeq 0$, $N_A^{1/3}\simeq A^{1/3}$ and $N_B^{1/3}\simeq B^{1/3}$; for very large $b$, $N_A^{1/3}\simeq N_B^{1/3}\simeq 1$. In general, $N_A^{1/3}$ and $N_B^{1/3}$ can be theoretically calculated or experimentally determined, by using forward/backward calorimeters to detect the non-interacting nucleons: $(A - N_A)$ in the forward direction and $(B- N_A^{2/3} N_B^{1/3})$ in the backward direction.

We distinguish again two regimes: 
\renewcommand{\theenumi}{\roman{enumi}}
\renewcommand{\labelenumi}{\theenumi)}
\begin{enumerate}
\item \textbf{High density regime} [$\Delta < \Delta_0$]\\
In this case all the strings (valence and sea) contribute, their numbers being proportional to the number of collisions
\begin{equation}
	\bar{N_s} \sim N_{coll}\sim N_A N_B^{1/3}\, ,
\end{equation}
and we obtain, from (\ref{eq:rhoperc}), 
\begin{equation}\label{eq:rhohigh}
	\rho(\Delta\sim-Y) \sim N_A^{5/6} N_B^{1/6}\, .
\end{equation}
\item \textbf{Low density regime} [$\Delta > \Delta_0$]\\
The number of strings is essentially proportional to  the number of valence strings, which is in turn proportional to the number of participating  --- forward or backward --- nucleons,
\begin{equation}
	\bar{N_s} (\Delta\sim 0) \sim N_{part}\,\,\mbox{(Forward/Backward)}\, ,
\end{equation}
We obtain, see (\ref{eq:rhoperc}),
\begin{equation}\label{eq:rholowforw}
	\rho(\Delta\sim 0) \sim N_A\,\,\mbox{(Forward)}\ \, ,
\end{equation}
and
\begin{equation}\label{eq:rholowback}
	\rho(\Delta\sim 0) \sim N_A^{2/3} N_B^{1/3}\,\,\mbox{(Backward)}  \, .
\end{equation}
\end{enumerate}

Bearing in mind (\ref{eq:rhonucleus}), (\ref{eq:rhohigh}), (\ref{eq:rholowforw}) and(\ref{eq:rholowback}) we arrive at
\begin{align}
	\rho( \Delta) &=  \frac{N_A^{5/6} N_B^{1/6} \rho_Y}{\Big(\frac{N_B}{N_A}\Big)^{1/6} e^{\frac{\Delta - \Delta_0}{\delta}}+1}\,\, \mbox{(Forward)}\, ,\label{eq:rhofinalforw}\\
	\rho( \Delta) &=  \frac{N_A^{5/6} N_B^{1/6} \rho_Y}{\Big(\frac{N_A}{N_B}\Big)^{1/6} e^{\frac{\Delta - \Delta_0}{\delta}}+1}\,\, \mbox{(Backward)}\, .\label{eq:rhofinalback}
\end{align}

If we impose, in (\ref{eq:rhofinalforw}) and (\ref{eq:rhofinalback}), the condition $N_B=N_A$ (i.e., a symmetric situation) we recover (\ref{eq:rhonucleus}) and the forward (\ref{eq:rhofinalforw}) and backward  (\ref{eq:rhofinalback}) distributions  become mirror images. Let us see, in this simple situation, how the forward and backward distributions can be matched.
The variable $\Delta$ starts, for $y=\pm Y$, at zero and decreases towards the mid-rapidity region reaching a minumum value. We thus have the constraint
\begin{equation}\label{eq:constraint}
	-\Delta^B_{MIN} - \Delta^F_{MIN} = 2 Y\, .
\end{equation}
In the symmetric case $\Delta^B_{MIN} = \Delta^F_{MIN}$ and 
\begin{equation}
	\Delta_{MIN} = - Y\, ,
\end{equation}
corresponding to $y=0$ and the distribution is centred around the centre of mass of the nucleon-nucleon system.

In the asymmetric case the denominators in (\ref{eq:rhofinalforw}) and (\ref{eq:rhofinalback}) are different, and the condition (\ref{eq:constraint}) is not sufficient to match the distributions. 
Further requiring
\begin{equation}\label{eq:constraint2}
	\Delta^B_{MIN} - \frac{\delta}{6} \ln \frac{N_B}{N_A}= \Delta^F_{MIN} +\frac{\delta}{6} \ln \frac{N_B}{N_A}\, ,
\end{equation}
we obtain, after combining (\ref{eq:constraint}) and (\ref{eq:constraint2}), 
\begin{align}
	- \Delta^F_{MIN} &= Y + \frac{\delta}{6} \ln \frac{N_B}{N_A}\, , \label{eq:matchf}\\
	- \Delta^B_{MIN} &= Y - \frac{\delta}{6} \ln \frac{N_B}{N_A}\, .\label{eq:matchb}
\end{align}
The conclusion is simple: the maximum of the distribution $\rho(\Delta)$ is displaced towards the backward region (higher nucleonic density) and for the same $\Delta$, $\rho_B(\Delta) > \rho_A(\Delta)$. These facts are qualitatively seen in the data \cite{Back:2004mr}.

We shall now be more specific regarding the quantities $\delta$, $\rho_Y$ and $\Delta_0$. We expect $\delta$ to not have a strong dependence on $Y$, being essentially a constant. The parameter $\rho_Y$ is the normalized particle density at mid-rapidity and thus should be related to the gluon distribution at small $x$, 
\begin{equation}\label{eq:gluonrise}
	\rho_Y \simeq e^{\lambda Y}\, ,
\end{equation}
with $\lambda \simeq 0.25 - 0.3$ \cite{Armesto:2004ud,Albacete:2005ef}. From energy-momentum conservation $\lambda$ was also estimated: $\lambda \simeq 2/7 - 1/3$ \cite{Brogueira:2006nz,DiasdeDeus:2005sq}.
Regarding $\Delta_0$ we cannot assume it to be constant, as in the Feynman-Wilson gas model. As $\rho_Y$ increases with rapidity $Y$ (\ref{eq:gluonrise}),   energy conservation arguments give that  $\Delta_0$ has to decrease linearly with $Y$. We write
\begin{equation}
	\label{eq:decdelta}
	\Delta_0 = -\alpha Y\, ,
\end{equation}
with $0<\alpha <1$.

Eq. (\ref{eq:rhonucleus}) can be rewritten as 
\begin{equation}\label{eq:rhoparam}
	\frac{1}{N_A} \rho (\Delta, Y) \equiv \frac{2}{N_{part}} \frac{dn}{dy} = \frac{e^{\lambda Y}}{e^{\frac{\Delta +\alpha Y}{\delta}} +1}\, .
\end{equation}
In fact, (\ref{eq:rhoparam}) was used in \cite{Adams:2005cy} as a \textit{suitable parametrization} of 
RHIC $Au-Au$ data from \cite{Back:2004je}.

\section{Limiting fragmentation}
\label{sec:limfrag}

Recently, the phenomenon of limiting fragmentation \cite{Benecke:1969sh}  has received renewed attention \cite{Brogueira:2006nz,Back:2004je,Jeon:2003nx,Jalilian-Marian:2002wq,Gelis:2006tb}.
The particle density $\rho$ is, in general, a function of both $\Delta$ and the beam rapidity $Y$. The limiting fragmentation hypothesis essentially states that for $\Delta$ larger than some $Y$-dependent threshold, the density $\rho$ becomes a function of $\Delta$ only.
Since $\Delta_0$ is a decreasing function of $Y$ (\ref{eq:decdelta}), limiting fragmentation can be recast, in the framework of our model, as
\begin{equation}
	\rho(\Delta, Y) \xrightarrow[Y\rightarrow \infty\, ,\,\,\Delta >\Delta_0]{} f(\Delta)\, .
\end{equation}

From (\ref{eq:rhoparam}) we can write
\begin{equation}
	\rho(\Delta, Y) \xrightarrow[Y\rightarrow \infty\, ,\,\,\Delta >\Delta_0]{} 
	e^{-\Delta/\delta} e^{(\lambda -\alpha/\delta) Y}\, ,
\end{equation}
so that limiting fragmentation will hold if, for $\delta >0$, one has
\begin{equation}\label{eq:fragcond}
	\frac{\alpha}{\lambda \delta} =1\, .
\end{equation}
RHIC data \cite{Back:2004je} does  not seem to favour (\ref{eq:fragcond}) as recently observed in \cite{Brogueira:2006nz}, the tendency being for the curve, as the energy increases, to approach a step function. 
However, the BRAHMS collaboration \cite{Bearden:2003hx} points out that the net baryon rapidity loss seems to saturate between SPS and RHIC energies, which certainly is an indication for the limiting fragmentation regime.
In our opinion, no strong statements should be made as the considered range in $Y$ is small and QCD corrections most certainly introduce some weak violation of limiting fragmentation. 

If we assume (\ref{eq:fragcond}) to hold, that is that limiting fragmentation should hold in some limit, we arrive at the conclusion that the evolution of $\rho (\Delta, Y)$ with  $Y$ is governed by the same evolution equation as its evolution with $\Delta$ (\ref{eq:log}).

In fact, from  (\ref{eq:rhoparam}) and (\ref{eq:fragcond}), one easily arrives at  
\begin{equation}
	\frac{1}{N_A} \rho (\Delta, Y) = \frac{e^{-\Delta/\delta}}{e^{-\lambda Y} e^{-\Delta/\delta} +1}\, ,
\end{equation}
which can be obtained from (\ref{eq:rhoparam}) through the change
\begin{equation}\label{eq:change}
	\lambda Y \longleftrightarrow -\Delta/\delta\, .
\end{equation}
Eq.~(\ref{eq:change}) means that the evolution equation in $Y$ is, see  (\ref{eq:log}) and (\ref{eq:rhos}), 
\begin{equation}
	\label{eq:yevol}
	\frac{\partial \rho (\Delta, Y)}{\partial Y} = \lambda \Big(\rho - \frac{1}{\rho_\Delta}\rho^2\Big)\, ,
\end{equation}
where $\rho_\Delta = e^{-\Delta/\delta}$ (obtained from $\rho_Y$ (\ref{eq:gluonrise}) by using (\ref{eq:change})) is the maximum value of $\rho$. 
If one rescales  $\rho$ to a function with maximum value $1$, i.e.  perform the transformation $\rho \longrightarrow \rho_\Delta\, \rho$, then (\ref{eq:yevol})  can be rewritten as
\begin{equation}
	\label{eq:bkanal}
	\frac{\partial \rho (\Delta, Y)}{\partial Y} = \lambda \rho \big(1- \rho\big)\, ,
\end{equation}
which is formally equivalent to the Balitsky-Kovchegov (BK) equation \cite{Balitsky:1995ub,Kovchegov:1999yj} for the $Y$ evolution of the scattering amplitude of a colour dipole once the dependence on transverse coordinates  has been neglected.

\section{Conclusions}
\label{sec:conc}

We explored the connection between the Glasma and an effective description in terms of the SPM by introducing a simple model for the distribution in centre of mass rapidity of the particle density.

The solution of this model for a symmetrical nucleus-nucleus collision (\ref{eq:rhonucleus}), i.e. a collision of identical nuclei, reproduces (\ref{eq:rhoparam}) the form used as a \textit{parametrization} of RHIC $Au-Au$ data \cite{Adams:2005cy}.

In the case of an asymmetrical collision the maximum of the distribution is, in qualitative agreement with RHIC data  \cite{Back:2004mr}, displaced towards backward (higher nucleonic density) region.
The expression obtained for the forward (\ref{eq:rhofinalforw}) and backward  (\ref{eq:rhofinalback}) distributions, supplemented by the matching conditions (\ref{eq:matchf}) and (\ref{eq:matchb}),  can be used, since it also reduces to (\ref{eq:rhoparam}) for the limiting case of a symmetrical collision,   to perform a comprehensive fit of RHIC data and to yield predictions for $Pb-Pb$ collisions at the LHC \cite{Abreu:2007zw}.

Our model yields, in terms of its parameters, a necessary condition (\ref{eq:fragcond}) for the occurrence of limiting fragmentation. 
This condition is not favoured by current RHIC data. The fit of RHIC data performed in \cite{Brogueira:2006nz} yields values for the parameters ($\lambda = 0.247$, $\alpha=0.269$ and $\delta= 0.67$) that do not fulfil  (\ref{eq:fragcond}).
Further, attempting to perform a fit with  (\ref{eq:fragcond}) as a constraint results in a poor description of the data.
It should, however, be noted that, in the fragmentation region, (\ref{eq:rhoparam}) with parameter values obtained without enforcing limiting fragmentation (as above) results in curves \cite{Brogueira:2006nz} that are indistinguishable from a universal behaviour.
Our model predicts a clear violation of this apparent universal behaviour in the fragmentation region for $Pb-Pb$ collisions at the LHC. 

To the best of our knowledge, (\ref{eq:fragcond}) first written in  \cite{Brogueira:2006nz} and obtained here from a physically motivated explicit model, is the only possible indication that the apparent universal behaviour of the fragmentation region in RHIC data may not correspond to limiting fragmentation.

If limiting fragmentation is enforced into our model, that is the fulfilment of (\ref{eq:fragcond}) is imposed, than an equation (\ref{eq:yevol}) for the evolution of the particle density with total rapidity $Y$ can be written. This equation can, in turn, be rescaled to an  equation  (\ref{eq:bkanal}) \textit{formally} identical to the BK equation with no dependence on transverse degrees of freedom.
A note of caution as to the scope of this analogy is due.
The equation obtained here  (\ref{eq:bkanal}) applies to the string/gluon field configuration existing immediately after the collision and describes the evolution with $Y$ of the multiplicity of strings (i.e., particle density) of any given rapidity length $\Delta$.
CGC inspired studies of limiting fragmentation \cite{Jalilian-Marian:2002wq,Gelis:2006tb} rely on that gluon production in the fragmentation region of each of the colliding nuclei can be described, at leading order, in the framework of $k_\perp$-factorization such that the inclusive cross section for gluon production can be written as the convolution of the unintegrated gluon distributions of the two nuclei.
The kinematical region of fragmentation is such that the relevant Bjorken-$x$ domains of the fragmenting nucleus (projectile) and the target are very different: large Bjorken-$x$ for the projectile and small Bjorken-$x$ for the target.
The BK equation is then used to evolve the target to small-$x$ (large $Y$), while the large-$x$ gluon distribution of the projectile is obtained from a phenomenological extrapolation \cite{Gelis:2006tb}. 
The BK dynamics guarantees that the typical transverse momentum in the projectile is much smaller than that in the target and that unitarity of the target evolution is preserved.
Limiting fragmentation follows naturally from the target reaching its black disk limit and the gluon production cross section being determined by the universal large-$x$ component of the projectile.

From the above discussion it should be clear that our formal analogue of BK  (\ref{eq:bkanal}) and the BK equation as used in CGC based studies of limiting fragmentation apply to different stages of the collision and should not be compared directly. If, however, in future works one finds an evolution equation with $Y$ for particle density distributions which is formally analogous to the BK equation with no transverse dependence, then this will trivially lead to limiting fragmentation. Whether such a strong statement can be made for formal analogues of the full BK equation or for analogues of equations beyond the mean field approximation of BK is not clear to us and clearly merits further consideration.

\section*{Acknowledgements}

The authors wish to thank N\'estor Armesto for many enlightening discussions and for a critical reading of a version of the manuscript.
J.~G.~M.  acknowledges the financial support of  the Funda\c c\~ao para a Ci\^encia e a Tecnologia  of Portugal (contract SFRH/BPD/12112/2003) and thanks the Galileo Galilei Institute for Theoretical Physics for the hospitality and INFN for partial support during the completion of this work.


\end{document}